\documentclass[11pt]{article}
\usepackage{cospar}
\usepackage{url}

\def\pdot {\dot P}
\def\ltsima{$\; \buildrel < \over \sim \;$}
\def\lsim{\lower.5ex\hbox{\ltsima}}

\usepackage{graphicx}
\usepackage[figuresright]{rotating}

\title{AXPS AND X-RAY DIM NEUTRON STARS: RECENT XMM-NEWTON AND CHANDRA RESULTS}

\author{F. Haberl\address{Max-Planck-Institut f\"ur extraterrestrische Physik, 
    Postfach 1312, 85741 Garching, Germany}}

\begin{document}

\maketitle

\begin{abstract}
To date more than half a dozen X-ray dim isolated neutron stars are
known. Their X-ray spectra are characterized by soft blackbody-like
emission (kT $\sim$ 50-120 eV) without indication for harder,
non-thermal components. These stars apparently show no radio
emission and no association with supernova remnants. Four of
them exhibit pulsations in their X-ray flux with periods in the
range of 8.39 s to 22.7 s. Similar pulse periods were found from
AXPs, maybe suggesting evolutionary connections. In this talk
I will review the recent observational results from X-ray dim
isolated neutron obtained by the major X-ray observatories
XMM-Newton and Chandra. Spectral and temporal characteristics,
inferred luminosities and magnetic field strengths will be
compared with those of AXPs.
\end{abstract}

\section*{INTRODUCTION}

ROSAT observations revealed the existence of a group of soft X-ray sources
with well defined and unique characteristics. To date we know seven or eight
such objects which are summarized in Table~1. Their energy spectra are well
modeled by blackbody emission with kT$_{BB}$ between 50 and 120 eV and no
significant flux changes are observed on time scales of years. No association 
with a supernova remnant was found and upper limits
derived from radio surveys indicate that the radio luminosity is well below  
that of ordinary radio pulsars (for recent reviews see Treves et al. 2000, 
Motch 2001). X-ray pulsations were discovered from two sources in ROSAT data
and recent Chandra and XMM-Newton observations identified two further stars
as pulsars. Low hydrogen column densities derived from the X-ray spectra
indicate that we deal with close and therefore low-luminosity objects (X-ray 
dim isolated neutron stars, XDINs).

Although it is now well established that the ROSAT sources are isolated
neutron stars, the origin of their soft X-ray emission
is not clear. The temperatures in the range of (0.6 -- 1.4)$\times 10^6$ K
inferred from their blackbody-like spectra are typical for young cooling 
neutron stars with ages less than 10$^6$ years. However these young 
rotation powered pulsars show radio emission and neutron star spin periods 
well below 1 s, both properties in contrast to that of the new ROSAT sources.
The long periods would require that these neutron stars were born with either
an unusually long period or an unusually strong magnetic field to decelerate
the rotation of the neutron star to the current value within its life time 
(Heyl and Hernquist, 1998). A decaying strong magnetic field could provide a 
significant source for the X-ray luminosity which, at 
least in the case of RX\,J0720.4--3125, is too high to be explained by rotational 
energy losses. Heating by magnetic field decay could also keep the high B field
pulsars detectable in X-rays over longer times than ordinary pulsars 
(Heyl and Kulkarni, 1998). 

Alternatively, old ($>10^9$ years) neutron stars could be re-heated by accretion
from the interstellar medium, as originally proposed by Ostriker, Rees and Silk 
(1970). Population synthesis studies by Treves and Colpi (1991), Blaes and Madau 
(1993) and Madau and Blaes (1994) estimated the number of old, isolated neutron 
stars detectable in the ROSAT all-sky survey to several thousands. Despite intensive
searches however, it became clear that the large galactic population of old neutron 
stars does not accrete from the interstellar medium at the originally predicted 
rates (Neuh\"auser and Tr\"umper, 1999; Motch 2001). Two possible explanations for
the elusiveness of old isolated neutron stars were elaborated by Treves et al. 
(2000). The velocity distribution may peak at $\sim$200--400 km s$^{-1}$, strongly
suppressing spherical accretion, or magnetic field decay on timescales of 10$^{8-9}$
years could prevent a large fraction of neutron stars from entering the accretor 
stage. The first evidence for a high space velocity of 185 km s$^{-1}$ was 
found for RX\,J1856.5--3754, the brightest and possibly nearest (117 $\pm$ 12 pc 
derived from parallax measurements) of the new isolated neutron stars (Walter and 
Lattimer, 2002).

A crucial neutron star parameter for both models is the magnetic field 
strength and its temporal evolution. The study of spin histories and measuring
spin-down rates can provide estimates for the magnetic field strength, when
interpreted in terms of magnetic dipole braking. First $\pdot$ measurements for
RX\,J0720.4--3125 yield a magnetic field strength of the order of 10$^{13}$ G 
(Kaplan et al., 2002a; Zane et al., 2002). Although, this value is below those 
typically quoted for magnetars (10$^{14-15}$ G) it is still extreme and close to
the critical value B$_c$ = 4.4$\times$10$^{13}$ G at which quantum effects 
become important. Anomalous X-ray pulsars (AXPs, for a recent review see 
Mereghetti et al., 2002) and soft $\gamma$-ray repeaters (SGRs, for a review
see Hurley, 2000), which share some common characteristics (Kaspi, 2003), may
belong to this class of neutron stars with the strongest known magnetic field
strengths. It is remarkable that the spin periods of AXPs, SGRs and XDINS cluster
within a narrow range. To further understand the properties of XDINs and their
stellar evolution it is therefore crucial to observe them with new generation
instruments available on board of Chandra and XMM-Newton. Recent results obtained
from high resolution spectroscopy and timing measurements are reviewed in the 
following.

\begin{table}
\caption{X-ray dim isolated neutron stars}
\begin{tabular}{lcccccll}

\noalign{\smallskip}\hline\noalign{\smallskip}
  Source   & $P$  & $\pdot$      & L$_{x}$       & kT$_{BB}$ & d     & Opt. & Comments \\
           & (s)  & (s~s$^{-1}$)   &(erg s$^{-1}$) & (eV)     &  (pc) & &  \\
\noalign{\smallskip}\hline\noalign{\smallskip}
         &         &  &                      &          & &  \\
RX\,J0420.0$-$5022 &    22.69    & $-$    & $2.7\times10^{30}$     & 57    & 100 & B$>$25.5 &  (1)  \\
\noalign{\smallskip}
RX\,J0720.4$-$3125  &  8.39     & $(3-6)\times10^{-14}$  & $2.6\times10^{31}$    &  85  & 100  & B=26.6 &  (2,3,4,5) \\
\noalign{\smallskip}
RX\,J0806.4$-$4123   &  11.37   & $-$ &        $5.7\times10^{30}$ & 95   & 100 &  B$>$24 & (6,7)  \\
\noalign{\smallskip}
1RXS\,J130848.6+212708 &   10.31   &  $< 6\times10^{-12}$ &   $5.1\times10^{30}$ & 90   & 100 & m$_{\rm 50CCD}$ & RBS 1223 \\
  &       &  &    &    & & =28.6 & (8,9,10)  \\
\noalign{\smallskip} 
RX\,J1605.3+3249   &   $-$    & $-$ &        $1.1\times10^{31}$ & 92   & 100 & B$>$27 & RBS 1556  (11) \\
 \noalign{\smallskip}
RX\,J1836.2+5925?   &   $-$    & $-$ &        $5.4\times10^{30}$ & 43   & 400 & V$>$25.2 & variable? (12) \\
RX\,J1856.5$-$3754   &   $-$      & $-$ &  $1.5\times10^{31}$  & 62 & 117 & V=25.7 & (13,14,15,16) \\
\noalign{\smallskip}
1RXS\,J214303.7+065419   &  $-$ & $-$ &   $1.1\times10^{31}$ & 90   & 100 & R$>$23 & RBS 1774  (17) \\
\noalign{\smallskip}\hline\noalign{\smallskip}
      \end{tabular}

(1) Haberl, et al., 1999;
(2) Haberl, et al., 1997;
(3) Paerels, et al., 2001;
(4) Cropper, et al., 2001;
(5) Zane, et al., 2002;
(6) Haberl, et al., 1998;
(7) Haberl \& Zavlin, 2002;
(8) Schwope, et al., 1999;
(9) Hambaryan, et al., 2002;
(10) Kaplan, et al., 2002b;
(11) Motch, et al., 1999;
(12) Mirabal, Halpern, 2001;
(13) Walter, et al., 1997;
(14) Walter \& Lattimer, 2002;
(15) Burwitz, et al., 2001;
(16) Drake, et al., 2002;
(17) Zampieri, et al., 2001.
\end{table}

\section*{MAIN PROPERTIES OF X-RAY DIM ISOLATED NEUTRON STARS}
\subsection*{Optical Identifications}

Optical counterparts were identified for the two brightest XDINs RX\,J1856.5--3754
(Walter and Matthews, 1997) and RX\,J0720.4--3125 (Motch and Haberl, 1998; Kulkarni 
and van Kerkwijk, 1998). Their X-ray to optical flux ratios are
log(f$_x$/f$_{opt}$) = 4.8 and
5.3, respectively. For 1RXS\,J130848.6+212708 Kaplan et al. (2002b) found a 
likely optical counterpart with log(f$_x$/f$_{opt}$) = 4.9 which would be even 
larger if the faint star is not the counterpart. For all other XDINS empty X-ray 
error circles yield lower limits for log(f$_x$/f$_{opt}$) in the range 2.5 for 
RX\,J1836.2+5925 to 4.0 in the case of RX\,J1605.3+3249 (see the discovery papers
of each individual source given as first reference in the comment column of Table~1).
These limits practically rule out any other possibility than an isolated neutron star.

XMM-Newton and in particular Chandra observations will allow to strongly reduce
the uncertainties in the X-ray positions, which is highly demanded to optical 
identification programs of such faint stars. An example is the likely identification 
of 1RXS\,J130848.6+212708 from above which is based on a 1 arcsec error radius obtained 
by Chandra. A first accurate X-ray position for RX\,J0806.4$-$4123 was recently reported
by Haberl et al. (2002).

\subsection*{X-ray Spectra}

The X-ray spectra obtained from most of the XDINs by the ROSAT PSPC were
consistent with little absorbed blackbody emission. However the low energy resolution
of the PSPC did not allow to look for absorption features which may be created in
a stellar atmosphere containing heavy chemical elements. However first high resolution 
spectra obtained by the RGS instruments on XMM-Newton of RX\,J0720.4$-$3125 
(Paerels et al., 2001) and by the LETGS on Chandra of RX\,J1856.5$-$3754 
(Burwitz et al., 2001, 2003) did not reveal any absorption features.
The absence of significant absorption features in the RGS spectra of 
RX\,J0720.4$-$3125 in the 0.35--1.25 keV range would appear to exclude magnetic fields of
B$\sim$(0.3--2.0)$\times$10$^{11}$ and (0.5--2.0)$\times$10$^{14}$ G which would cause 
electron or proton cyclotron resonances, respectively, in the observed energy band
(cf. Pavlov et al., 1995; Zane et al., 2000). Similarly, the LETGS spectrum (0.15--0.82 keV)
would exclude B$\sim$(1.3--7.0)$\times$10$^{11}$ and (0.2--1.3)$\times$10$^{14}$ G for
RX\,J1856.5$-$3754. Remarkably the X-ray spectra from both XDINs are best fit with a simple 
blackbody model. 

For some of the fainter XDINs medium resolution CCD spectra are available. RX\,J0806.4$-$4123
was observed by the EPIC instruments on XMM-Newton (Haberl and Zavlin, 2002) and Chandra 
ACIS-S spectra were published for 1RXS\,J130848.6+212708 by Hambaryan et al. (2002). 
For both sources the spectra are well modeled by blackbody spectra. Temperatures included 
in Table~1 are those obtained from Chandra and XMM-Newton when available (for RX\,J0720.4$-$3125, 
RX\,J0806.4$-$4123, 1RXS\,J130848.6+212708 and RX\,J1856.5$-$3754).

\subsection*{Spectral energy distribution}

The optical fluxes measured from the counterparts of XDINs
are higher than one would expect from the extrapolation
of the X-ray spectrum. Factors around 5 are derived (which however depend on the 
parameters used for the X-ray spectrum which are not always consistent between
different instruments) for RX\,J1856.5$-$3754 (Pons et al., 2002; Burwitz et
al., 2003) and RX\,J0720.4$-$3125 (Motch and Haberl, 1998). A similar value 
is found for the candidate counterpart of 1RXS\,J130848.6+212708 (Kaplan 
et al., 2002b). 

Pavlov et al. (1996) and Pons et al. (2002) have demonstrated that the light element
(hydrogen and helium) non-magnetic neutron star atmosphere models can not
reproduce the observed energy distribution of RX\,J1856.5$-$3754 from the 
optical to the X-ray band, as they over-predict the optical flux by two orders
of magnitude. Heavy element (iron and standard solar-mixture) atmosphere 
models (Zavlin and Pavlov, 2002) for reasonable values for the gravitational 
redshift do not fit the X-ray spectrum of RX\,J1856.5$-$3754 either (Burwitz et 
al., 2001). This let Pons et al. (2002) to propose a two component model
to explain the energy distribution of RX\,J1856.5$-$3754: a blackbody component
with kT = 55.3 eV and small area (R$_\infty$ = 8.2 km for a distance of 117 pc)
to mainly account for the X-ray emission and another blackbody component 
with kT = 20 eV and larger area (R$_\infty$ = 15.6 km) to match the observed 
optical flux. 

A problem with this hot polar cap interpretation is however (unless the spin axis
is either parallel to the magnetic axis or the observers line of sight) the
non-detection of X-ray pulsations from RX\,J1856.5$-$3754.
From the analysis of 505 ks Chandra data Ransom et al. (2002) and Drake et al.
(2002) derived upper limits of 4.5\% (10 ms -- 10$^3$ s) and 2.7\% (10 ms -- 
10$^4$ s), respectively. This limit was further reduced to 1.3\% 
(20 ms -- 10$^3$ s) by Burwitz et al. (2003) from a 57 ks XMM-Newton 
observation. In an attempt to explain this small variability, these authors 
discuss a one-component model with kT = 63 eV and R$_\infty$ = 12.0 km 
(distance 117 pc). To match optical and X-ray data, this model requires a low
radiative efficiency at X-ray energies, which may be expected if the neutron
star has a condensed matter surface (Zane 2003).

\subsection*{X-ray pulsations}

X-ray pulsations were discovered from two XDINs in ROSAT data: a 8.39 s period
for RX\,J0720.4--3125 (Haberl et al. 1997) and a possible period of 22.7 s
(4$\sigma$) from RX\,J0420.0$-$5022 (see Fig.~1). Hambaryan et al. (2002) 
reported the discovery of 
a 5.16 s modulation in the X-ray flux of 1RXS\,J130848.6+212708 observed by Chandra
ACIS-S. XMM-Newton observed this XDIN in Dec. 2001 for 20 ks (Hambaryan et al. in 
preparation). The derived power spectrum is shown in Fig.~2. Besides the strongest peak at 
the previously determined period of 5.16 s another peak at half the frequency suggests
that the true spin period of the neutron star is 10.31 s. The light curve folded at the
period of 10.31 s is shown in Fig.~1. Observed pulsed fractions are between 6\% 
(RX\,J0806.4$-$4123) and 43$\pm$14\% (RX\,J0420.0$-$5022). 
In the case of XDINs being cooling neutron stars the strongest spin modulations are difficult
to explain with temperature variations as function of magnetic colatitude which would be 
caused by e.g. a thermal conductivity dependence on the magnetic field vector (see Cropper et al.,
2001 and references therein).

\begin{figure}
\hbox{
\includegraphics[width=66mm,angle=-90]{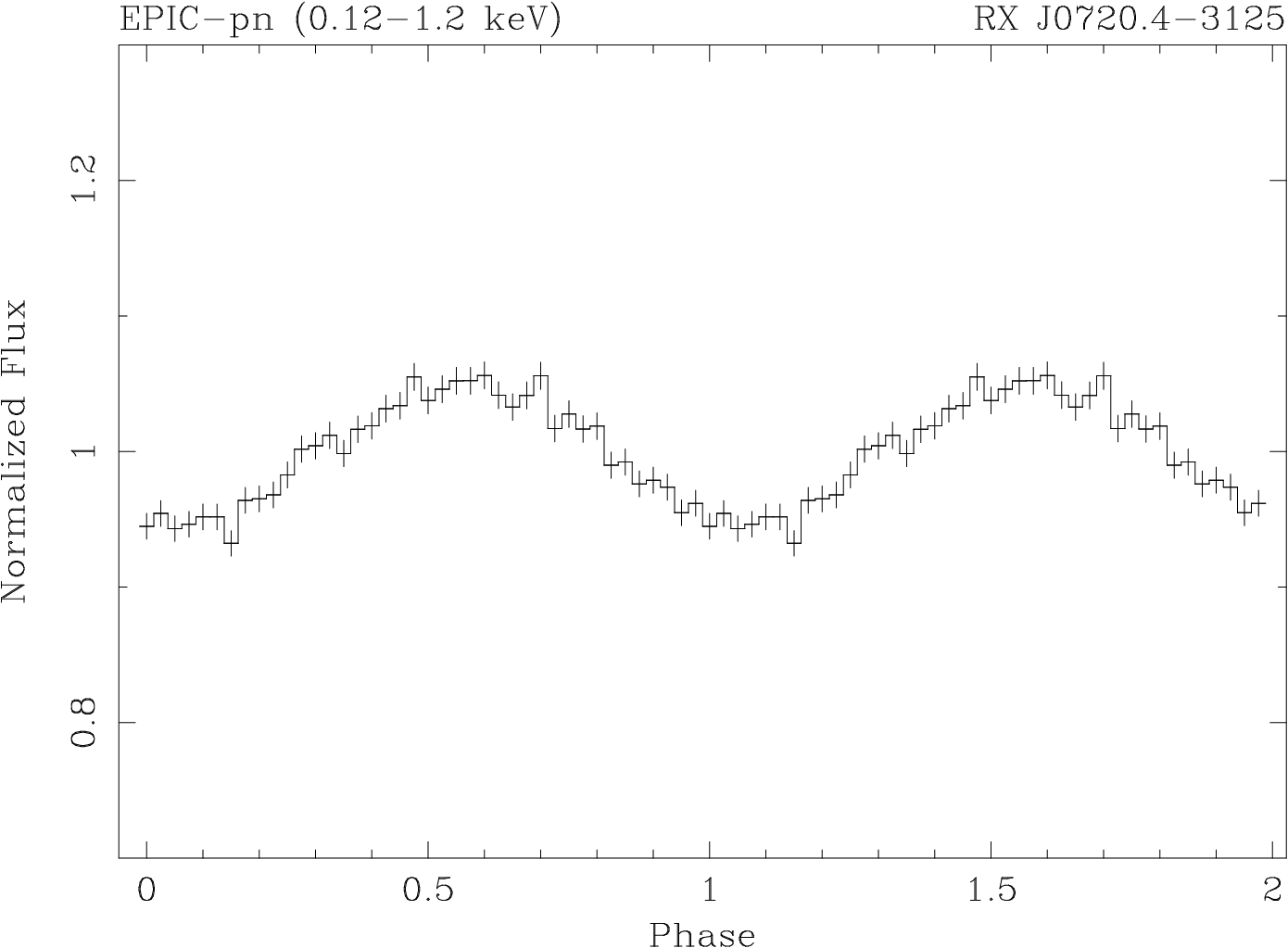}
\includegraphics[width=66mm,angle=-90]{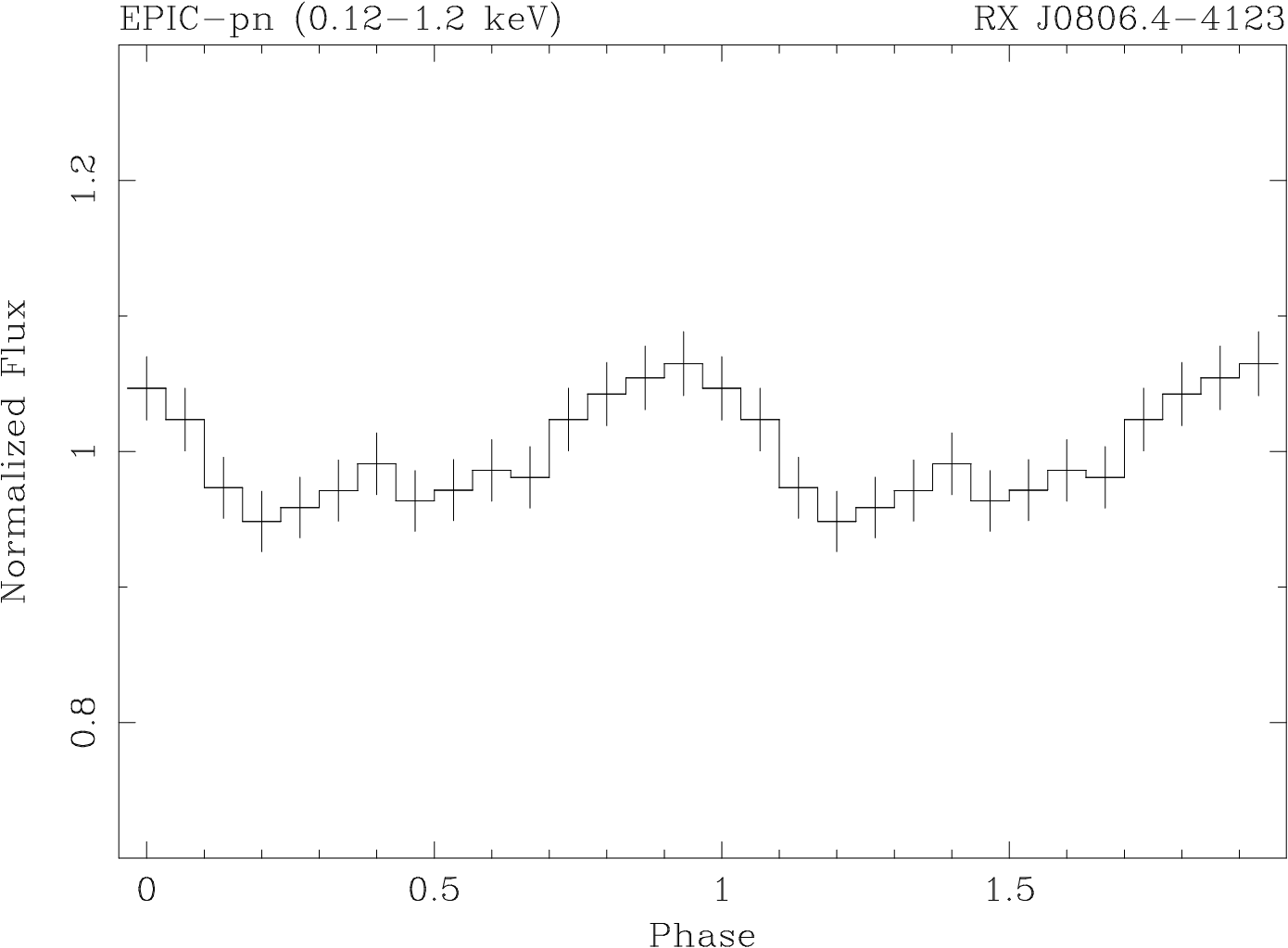}
}
\hbox{
\includegraphics[width=66mm,angle=-90]{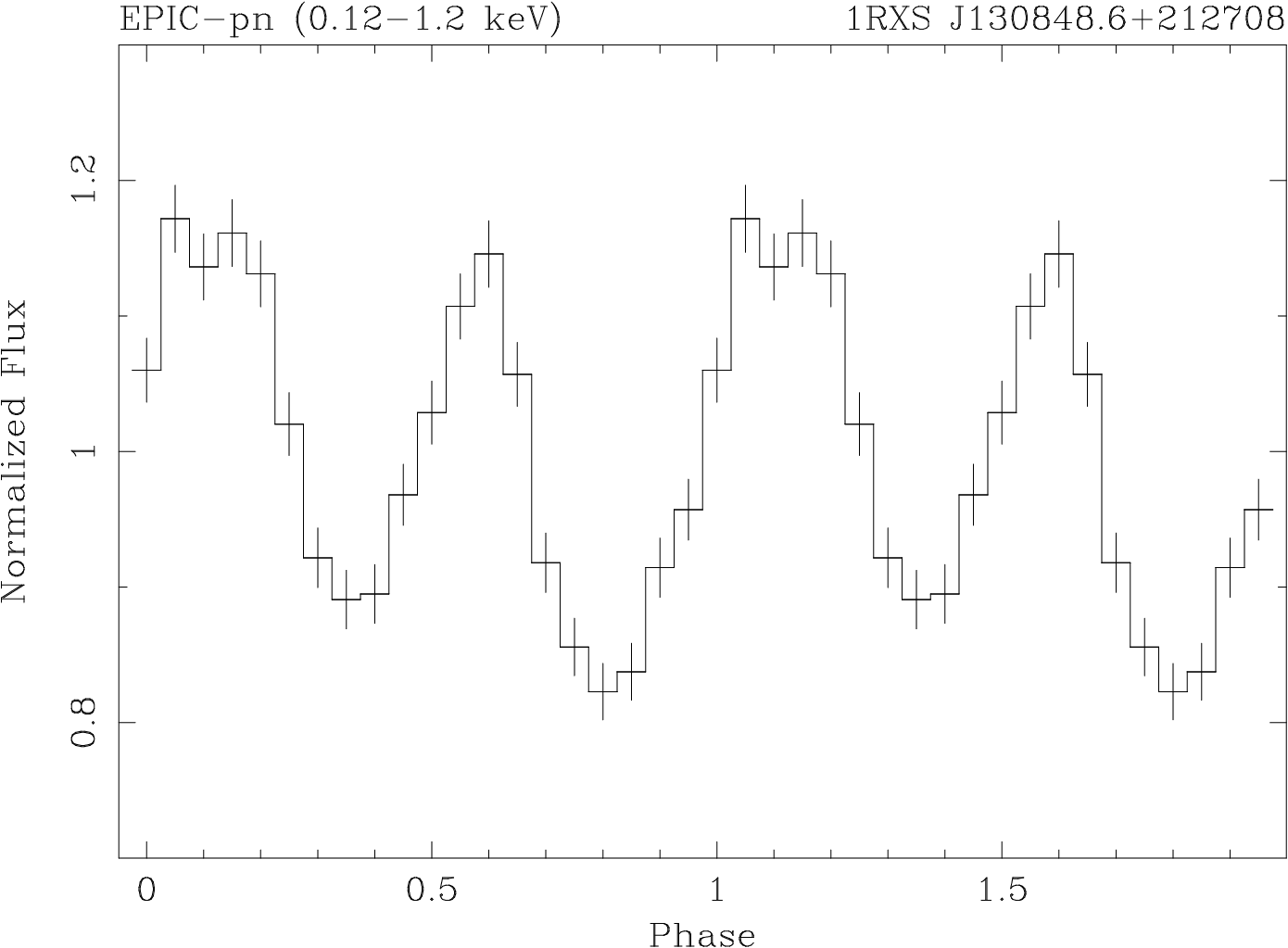}
\includegraphics[width=66mm,angle=-90]{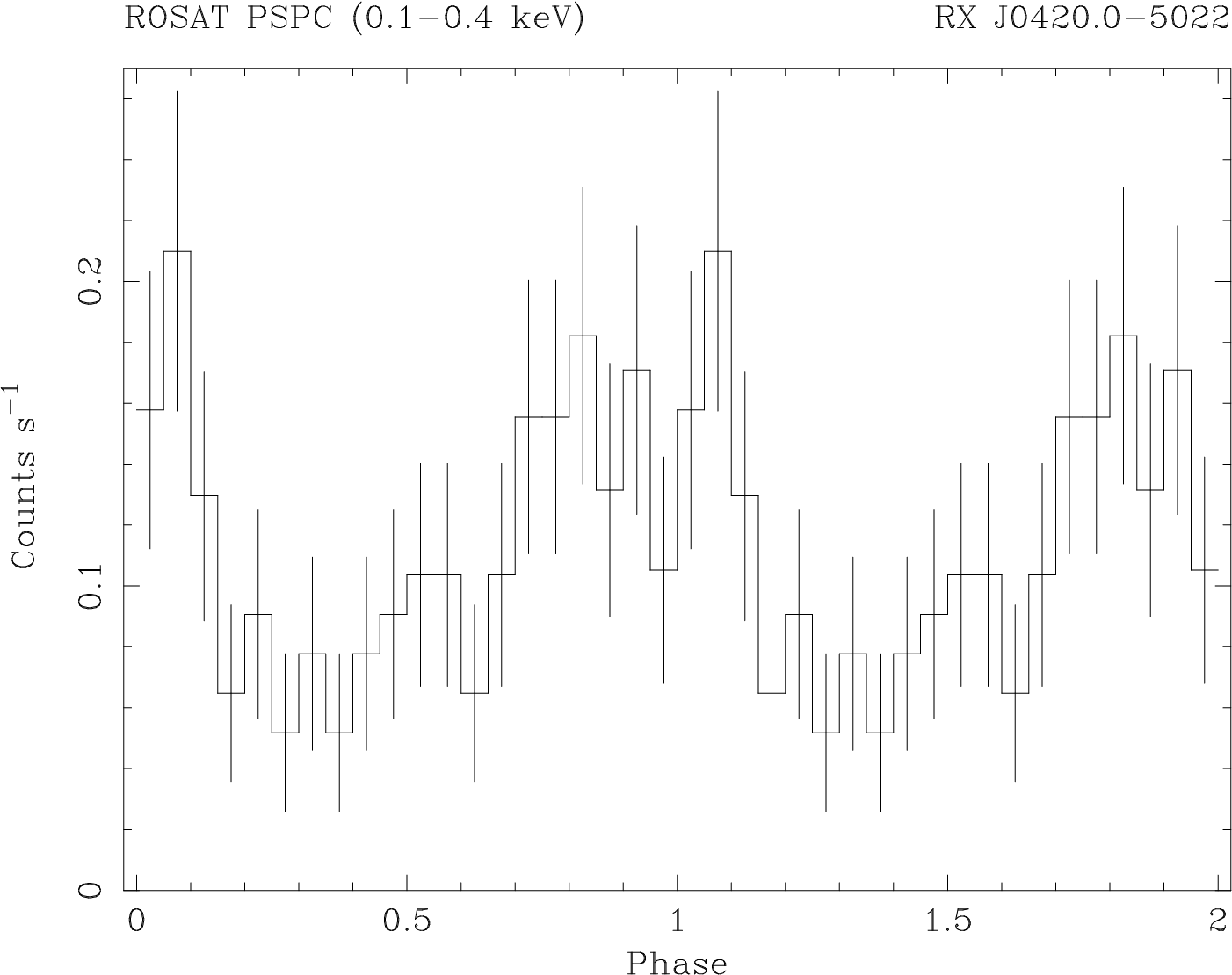}
}
\caption{Folded X-ray light curves of XDINs. For clarity two cycles are drawn. The three data 
sets obtained from XMM-Newton EPIC pn data were derived in a consistent way and are plotted 
on the same scale. For spin periods see Table~1.}
\end{figure}

\begin{figure}
\includegraphics[width=66mm,angle=-90]{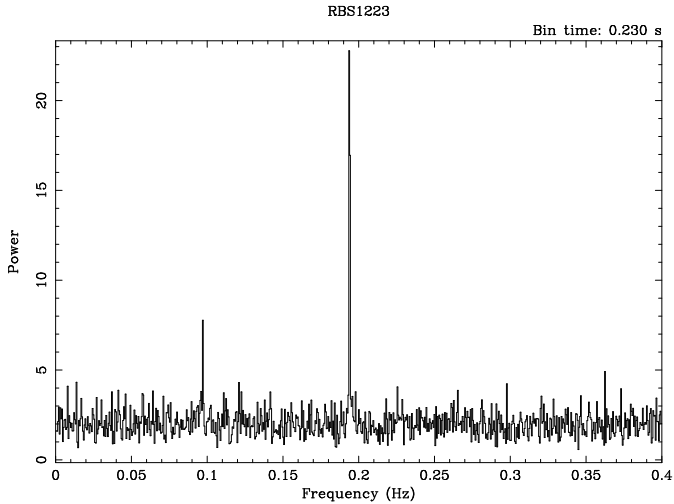}
\caption{Power spectrum of 1RXS\,J130848.6+212708 obtained from XMM-Newton EPIC-pn data.
A peak at half the known frequency indicates a double peaked pulse profile and a spin period of
10.31 s.}
\end{figure}

\subsection*{Spectral Variations}

Spectral variations with pulse phase were discovered from the 8.4 s pulsar RX\,J0720.4$-$3125
(Cropper et al., 2001). The (0.1--0.4 keV)/(0.4--0.8 keV) hardness ratio varies sinusoidal
similar to the pulsations but is softest slightly before flux maximum. Spectral fits to phase 
resolved spectra suggest that the majority of the variation is caused by a change in absorption
rather than a change in temperature (Cropper et al., 2001). While direct changes in photo-electric 
absorption might be conceivable in terms of material co-rotating in the magnetosphere, the 
hardness ratio modulation is more likely explained by energy dependent beaming effects. If 
such an energy dependence is similar to that of photo-electric absorption fitting an absorbed 
blackbody model would necessarily lead to variations in absorption. E.g. Burwitz et al. (2003) 
modeled the LETG spectrum of RX\,J1856.5$-$3754 with a blackbody modified by an energy dependent 
absorption factor of the form $\alpha$ = E$^\beta$ in addition to photo-electric absorption. 
They found a best fit with $\beta$=1.28$\pm$0.30 and a reduced column density. It could well be 
that in the case of RX\,J0720.4$-$3125 the spectral variation is caused by a phase dependent 
absorption factor, as it may be expected for a condensed matter surface, with varying viewing angles
relative to the magnetic field. Such models would not require large temperature variations 
on the surface of the neutron stars and greatly reduce the related problems in explaining
large pulsed fractions.

\subsection*{Spin Histories}

To measure spin period changes with time are of great importance to independently derive
estimates for the magnetic field strength of the neutron star. In particular in the
absence of absorption features in the high resolution spectra of XDINs it may be the 
only way to obtain this information. Measuring pulse period changes should also allow to 
discriminate between
models of cooling and accreting XDINs, with more erratic changes expected for the latter.
The best studied case is RX\,J0720.4$-$3125, the brightest XDIN showing X-ray pulsations.
Kaplan et al. (2002a) derived the first useful upper limit for the period derivative
of 3.6$\times 10^{-13}$ s s$^{-1}$ (3$\sigma$) from the analysis of ROSAT, SAX and Chandra
data. Including XMM-Newton data Zane et al. (2002) derived possible values of
3--6$\times 10^{-14}$ s s$^{-1}$  which gives a magnetic field strength of 2$\times 10^{13}$ G
when interpreted in terms of dipolar losses. This yields a dynamic age
(P/(2$\pdot$)) of 
$\sim$2.5$\times$10$^6$ years consistent with the age expected from theoretical cooling curves
for the measured temperature. Magnetic field decay may therefore not be required to explain the 
observed properties of RX\,J0720.4$-$3125.

A first value for $\pdot$ of (0.7--2.0) $\times 10^{-11}$ s s$^{-1}$ was measured for 
1RXS\,J130848.6+212708 (Hambaryan et al. 2002) relying on one period measurement from Chandra
and one uncertain (due to possible aliases) ROSAT value. The new period obtained from XMM-Newton 
data is however shorter than the Chandra value and a preliminary linear fit to the three 
available periods 
(using values twice as long as the originally reported value of 5.16 s) yields only an upper 
limit of $\pdot < 6 \times 10^{-12}$ s s$^{-1}$ leaving an indefinite situation for this XDIN.
A combined analysis of the data to explore the P and $\pdot$ parameter space is currently in 
progress (Hambaryan et al., in preparation).

\section*{DISCUSSION}

The new results obtained from optical and X-ray observations more and more support
the model of young cooling neutron stars for XDINs. The large proper
motion of RX\,J1856.5$-$3754 yields a small cross section for accretion from the 
interstellar matter and makes the X-ray luminosity difficult to explain.
Also the steady and low spin down rate observed from RX\,J0720.4$-$3125 and the infered 
high magnetic field strength is incompatible with accretion which would require
B $<$ 10$^{10}$ G (Haberl et al., 1997).

The properties of AXPs are reviewed in a recent article by Mereghetti et al.
(2002). The pulsating XDINs show similar pulse periods and also pulse profiles
as AXPs. Figure~3 shows the period distribution for both classes of objects.
In the light of the new period of 10.31 s for 1RXS\,J130848.6+212708
there may be some evidence for somewhat longer periods observed from XDINs on
average. However the number of known objects, in particular pulsating XDINS, is still
too low to draw definite conclusions. 
The 10.31 s period of 1RXS\,J130848.6+212708 implies a double-peaked pulse
profile which is similar to that of the AXPs 1E\,2259+586 and 4U\,0142+61.
However the latter changes the relative intensity of the two pulses at energies
above 4 keV, a property which may be related to the hard
power-law component in the spectra of AXPs which is probably not existent in 
XDINs. It is worth noting here, that like many other classes
of X-ray pulsars also XDINs apparently can exhibit double-peaked profiles,
indicating asymmetries between the two emitting areas.

\begin{figure}
\includegraphics[width=66mm,angle=-90]{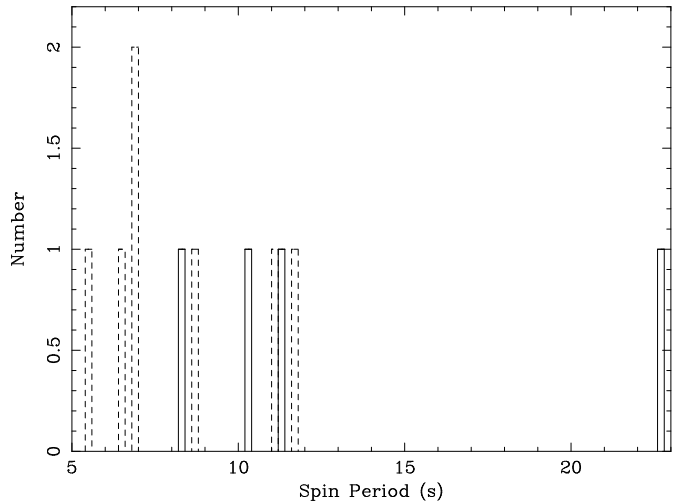}
\caption{Spin period distribution of AXPs (dashed) and XDINs.}
\end{figure}

The X-ray spectra of AXPs consist of a blackbody component with typical
temperatures around 400 to 600 eV, much hotter than those seen in XDINs
of 50 to 120 eV. Further, no hard emission component similar to the power-law
seen from AXPs has been detected yet from any of the XDINs. This may indicate
that AXPs are younger neutron stars (which is also supported by associations 
of some of them with supernova remnants) with a larger energy reservoir. The
currently favored model for AXPs is that of magnetars, neutron stars with very
strong magnetic fields of 10$^{14-15}$ G and the decay of the field providing
the energy. The question is now how do such magnetars look like after their
magnetic field has decayed to $\sim 10^{13}$ G, the value observed from
RX\,J0720.4$-$3125. In other words are XDINs descendents of AXPs?

XDINs are far more numerous as we can detect them only in a small volume
around the Sun (of a few hundred pc radius). The X-ray luminosities of 
10$^{34-36}$ erg s$^{-1}$ for AXPs allow their detection within the whole Galaxy.
From volume considerations maybe 10$^{3-4}$ more XDINs than AXPs currently 
exist in the Galaxy, a number however highly uncertain because it is not clear
how representative the solar neighborhood is in terms of volume density of
XDINs. A factor of about 100 (assuming evolutionary time scales of 10$^6$ and
10$^4$ years for XDINs and AXPs, respectively, during which we can detect
these sources) could be explained by evolution of AXPs into
XDINs. Hence, only 1-10\% of XDINs may be descendents of AXPs, the majority
being born with weaker (10$^{13}$ G) magnetic fields. One such case may be 
RX\,J0720.4$-$3125, for which no significant magnetic field decay is required
to explain its observed spin period history and temperature. The existence of 
such objects is naturally expected if one accepts the idea of magnetars with
magnetic field strengths above $\sim 10^{14}$ G and a field strength
distribution at neutron star birth down to the 10$^{12}$ G typically observed
from X-ray binary pulsars. 

If AXPs cool down during their evolution, loose their X-ray activity and become 
XDINs, an important question is also why we do not see any isolated neutron
star with intermediate temperature of 100--400 eV. MS\,0317.7$-$6647 was 
the first object proposed for an old accreting neutron star on the basis of its soft X-ray
spectrum: Stocke et al. (1995) presented a ROSAT PSPC spectrum and derived a 
blackbody temperature of 180 eV. The source is located in the field of the
nearby galaxy NGC\,1313 (and is also called NGC\,1313 X-2) and an interpretation 
as ultra-luminous X-ray source in NGC\,1313 could not be excluded. This 
latter possibility is now strongly supported by 
XMM-Newton observations of NGC\,1313 which reveal an additional hard component
in the spectrum of X-2 and its similarity to X-1, another ultra-luminous X-ray source in 
NGC\,1313 (Miller et al., 2003). 
Therefore, the ROSAT discovered XDINs are probably the only XDINs we now today.
However, it should be remarked here, that their low temperatures might partly be due to a 
selection effect. Searches in ROSAT all-sky survey data were performed among the very
softest sources (Haberl et al., 1998) and the simply larger number of harder ROSAT sources
makes the identification as XDIN more difficult.
Evolutionary scenarios remain therefore speculation until more
information about magnetic field strength of XDINs is obtained and more of these
enigmatic objects are discovered.


\end{document}